\documentclass[a4paper,preprint]{emulateapj}
\usepackage{amstext}
\usepackage{apjfonts}
\usepackage{amsmath}

\def\gsim{ \lower .75ex \hbox{$\sim$} \llap{\raise .27ex \hbox{$>$}} }
\def\lsim{ \lower .75ex\hbox{$\sim$} \llap{\raise .27ex \hbox{$<$}} }

\begin{document}

\title{Constraining the structure of GRB Jets
Through the $\log N - \log S$ Distribution}

\author{Dafne Guetta\altaffilmark{1,}\altaffilmark{2},
 Jonathan Granot\altaffilmark{3,4}
and Mitchell C. Begelman\altaffilmark{2}}

\altaffiltext{1}{Racah Institute for Physics, The Hebrew University,
  Jerusalem 91904, Israel; dafne@arcetri.astro.it}
\altaffiltext{2}{JILA, 440 UCB, University of Colorado, Boulder, CO
  80309; mitch@jila.colorado.edu}
\altaffiltext{3}{Institute for Advanced Study, Olden Lane, Princeton,
  NJ 08540}
\altaffiltext{4}{KIPAC, Stanford University, P.O. Box 20450, MS29,
Stanford, CA 94309; granot@slac.stanford.edu}

\begin{abstract}

A general formalism is developed for calculating the luminosity
function and the expected number $N$ of observed GRBs above a peak
photon flux $S$ for an arbitrary GRB jet structure. This formalism
directly provides the true GRB rate for any jet model,
instead of first calculating the GRB rate assuming isotropic
emission and then introducing a `correction factor' to account for
effects of the GRB jet structure, as was done in previous works.
We apply it to the uniform jet (UJ) and universal structured jet
(USJ) models for the structure of GRB jets and perform fits to the
observed $\log N-\log S$ distribution from the GUSBAD catalog
which contains 2204 BATSE bursts. We allow for a scatter in the
peak luminosity $L$ for a given jet half-opening angle $\theta_j$
(viewing angle $\theta_{\rm obs}$) in the UJ (USJ) model, which is
implied by observations. A core angle $\theta_c$ and an outer edge
at $\theta_{\rm max}$ are introduced for the structured jet, and a
finite range of opening angles $\theta_{\rm
min}\leq\theta_j\leq\theta_{\rm max}$ is assumed for the uniform
jets. The efficiency for producing $\gamma$-rays,
$\epsilon_\gamma$, and the energy per solid angle in the jet,
$\epsilon$, are allowed to vary with $\theta_j$ ($\theta_{\rm
obs}$) in the UJ (USJ) model, $\epsilon_\gamma\propto\theta^{-b}$
and $\epsilon\propto\theta^{-a}$. We find that a single power-law
luminosity function provides a good fit to the data. Such a
luminosity function arises naturally in the USJ model, while in
the UJ model it implies a power-law probability distribution for
$\theta_j$, $P(\theta_j)\propto\theta_j^{-q}$. The value of $q$
cannot be directly determined from the fit to the observed $\log
N-\log S$ distribution, and an additional assumption on the value
of $a$ or $b$ is required. Alternatively, an independent estimate
of the true GRB rate would enable one to determine $a$, $b$ and
$q$. The implied values of $\theta_c$ (or $\theta_{\rm min}$) and
$\theta_{\rm max}$ are close to the current observational limits.
The true GRB rate for the USJ model is found to be $R_{\rm
GRB}(z=0)=0.86^{+0.14}_{-0.05}\;{\rm Gpc^{-3}\;yr^{-1}}$
($1\;\sigma$) while for the UJ model it is higher by a factor
$f(q)$ which strongly depends on the unknown value of $q$.

\end{abstract}

\section{Introduction}

There are several lines of evidence in favor of jets in gamma-ray
bursts (GRBs). For GRBs with known redshift $z$, the fluence can
be used to determine the total energy output in $\gamma$-rays
assuming spherical symmetry, $E_{\rm \gamma,iso}$. The values of
$E_{\rm \gamma,iso}$ that were inferred in this way sometimes
approached, and in one case (GRB 991023) even exceeded $M_\odot
c^2$. Such high energies are very hard to reconcile with
progenitor models involving stellar mass objects. A non-spherical,
collimated outflow (i.e. a jet) can significantly reduce the total
energy output in $\gamma$-rays compared to $E_{\rm \gamma,iso}$,
since in this case the $\gamma$-rays are emitted only into a small
fraction, $f_b\ll 1$, of the total solid angle. A more direct (and
probably the best so far) line of evidence in favor of jets in
GRBs is from achromatic breaks in the afterglow light curves
\citep{Rhoads97,Rhoads99,SPH99}.

Despite the large progress in GRB research since the discovery of
the afterglow emission in early 1997, the structure of GRB jets is
still an open question. This is a particularly interesting and
important question, since it affects the total energy output and
event rate of GRBs, as well as the requirements from the central
engine that accelerates and collimates these relativistic jets.

The leading models for the jet structure are: (1) the uniform jet
(UJ) model
\citep{Rhoads97,Rhoads99,PM99,SPH99,KP00,MSB00,Granot01,Granot02},
where the energy per solid angle $\epsilon$ and the initial Lorentz
factor $\Gamma_0$ are uniform within some finite half-opening angle
$\theta_j$, and sharply drop outside of $\theta_j$, and (2) the
universal structured jet (USJ) model \citep{LPP01,RLR02,ZM02}, where
$\epsilon$ and $\Gamma_0$ vary smoothly with the angle $\theta$ from
the jet symmetry axis. In the UJ model the different values of the
jet break time $t_j$ in the afterglow light curve arise mainly due
to different $\theta_j$ (and to a lesser extent due to different
ambient densities). In the USJ model, all GRB jets are intrinsically
identical, and the different values of $t_j$ arise mainly due to
different viewing angles $\theta_{\rm obs}$ from the jet axis [in
fact, the expression for $t_j$ is similar to that for a uniform jet
with $\epsilon\to\epsilon(\theta_{\rm obs})$ and
$\theta_j\to\theta_{\rm obs}$]. The observed correlation $t_j\propto
E_{\rm \gamma,iso}^{-1}$ \citep{Frail01,BFK03} implies a roughly
constant true energy $E$ between different GRB jets in the UJ model,
and $\epsilon\propto\theta^{-2}$ outside of some core angle
$\theta_c$ in the USJ model \citep{RLR02,ZM02}.\footnote{The latter
is obtained assuming that the efficiency in producing $\gamma$-rays,
$\epsilon_\gamma$, does not depend on $\theta$. We later examine the
consequences of relaxing this assumption.} The probability
distribution of jet half-opening angles between different GRBs in
the UJ model, $P(\theta_j)$, is not given a-priori by the model, and
is a free function to be determined by observations.

Several methods have been used so far in order to constrain the
structure of GRB jets and help distinguish between the UJ model
and the USJ model. The afterglow light curves are similar for the
UJ and USJ models, but nevertheless some differences still exist
which might help distinguish between them \citep{KG03,GK03}.
Afterglow light curves also constrain the jet structure in the USJ
model, as discussed in \S \ref{lum_fun}. This method requires a
very good and dense monitoring of the afterglow light curves,
especially near the jet break time $t_j$. Another possible way to
distinguish between these two models for the jet structure is
through the different expected evolution of the linear
polarization during the afterglow \citep{Rossi04}. However, some
difficulties and complications exist in this method, like the poor
quality of most polarization light curves, and a possible ordered
magnetic field component that might cause polarization that is not
related to the jet structure \citep{GK03b}. The distribution of
viewing angles $\theta_{\rm obs}$, inferred from the observed
values of $t_j$, has been used to argue in favor of the USJ model
\citep{PSF03}. However, when the known redshifts of the same
sample of GRBs are also taken into account, there is a very poor
agreement with the predictions of the USJ model \citep{NGG04}. In
order to reach strong conclusions using this method, a large and
uniform sample of GRBs with known redshifts is needed, like the
one expected from {\it Swift}.

In this paper we use the observed $\log N -\log S$ distribution
(the number $N$ of GRBs above a limiting peak photon flux $S$) in
order to constrain the jet structure. A somewhat similar analysis
was done by \citet{GPW04}. They found that the observed $\log N
-\log S$ distribution rules out the USJ model, due to the paucity
of GRBs with small peak fluxes compared to the prediction of the
USJ model, and fit a double power-law luminosity function for the
UJ model. \citet{Firmani04} tried to constrain the redshift
evolution of the GRB rate and luminosity function, using the $\log
N -\log S$ distribution (as well as the redshift distribution
derived from the luminosity-variability relation).

This paper improves upon previous works by: (1) allowing the
efficiency in producing $\gamma$-rays to be a function of the
angle $\theta$, $\epsilon_\gamma=\epsilon_\gamma(\theta)$, and
varying $\epsilon(\theta)$ accordingly, using the \citet{Frail01}
relation, (2) including an internal dispersion in the peak
isotropic equivalent luminosity $L$ at any given angle $\theta$,
(3) introducing an inner core angle $\theta_c$ and an outer edge
$\theta_{\rm max}$ in the USJ model, (4) using a larger and more
uniform GRB sample.

In \S \ref{efficiency} we allow $\epsilon_\gamma$ and $\epsilon$
to vary with $\theta$, and derive the constraints on the
power-law indexes that are implied by the \citet{Frail01} relation. The
formalism for calculating the luminosity function and the observed
GRB rate as a function of the limiting flux for different jet
structures is derived in \S \ref{lum_fun}. In \S \ref{USJ} and \S
\ref{UJ} we compare the observed $\log N - \log S$ distribution to
the predictions of the USJ and UJ models, respectively. Our
results are discussed in \S \ref{discussion}.

\section{The Energy and Gamma-Ray Efficiency Distributions}
\label{efficiency}

The efficiency for producing $\gamma$-rays, $\epsilon_\gamma$, is
taken to be a function of the angle $\theta$ from the jet symmetry
axis in the USJ model, and a function of the jet half-opening
angle $\theta_j$ in the UJ model. For convenience we assume a
power-law dependence on $\theta$ in the USJ model,
$\epsilon_\gamma(\theta)=\Theta(\theta_{\rm
max}-\theta)\epsilon_{\gamma,0}\min[1,(\theta/\theta_c)^{-b_{\rm
USJ}}]$, where $\Theta(x)$ is the Heaviside step function. An
outer edge at $\theta_{\rm max}$ has been introduced, as well as a
core angle, $\theta_c\gtrsim 1/\Gamma_{\rm max}\sim 5\times
10^{-4}\;$rad, which is needed in order to avoid a divergence at
$\theta=0$. Here $\Gamma_{\rm max}\sim 2000$ is the maximum value
of the Lorentz factor to which the fireball can be accelerated
\citep{GSW01}. For the UJ model $P(\theta_j)$ is restricted to a
finite range of values, $\theta_{\rm
min}\leq\theta_j\leq\theta_{\rm max}$, and in analogy to the USJ
model we chose $\epsilon_\gamma(\theta_j)=\Theta(\theta_{\rm
max}-\theta_j)\Theta(\theta_j-\theta_{\rm
min})\epsilon_{\gamma,0}(\theta_j/\theta_{\rm min})^{-b_{\rm
UJ}}$.

The energy per solid angle is also assumed to behave as a power
law, $\epsilon(\theta)=\Theta(\theta_{\rm
max}-\theta)\epsilon_0\min[1,(\theta/\theta_c)^{-a_{\rm USJ}}]$ in
the USJ model and $\epsilon(\theta)=\Theta(\theta_{\rm
max}-\theta_j)\Theta(\theta_j-\theta_{\rm min})\epsilon_0\
(\theta_j/\theta_{\rm min})^{-a_{\rm UJ}}$ in the UJ model. The
power-law indexes $a$ and $b$ can be different between the USJ and
UJ models (as is emphasized by the different subscript for
the two models), and we also consider different values for $a$
and $b$ within each model.

The external mass density is taken to be a power-law in the
distance $R$ from the central source, $\rho_{\rm ext} = A R^{-k}$.
For a constant efficiency $\epsilon_\gamma$ ($b=0$) the observed
correlation $t_j\propto E_{\gamma,{\rm iso}}^{-1}$
\citep{Frail01,BFK03} implies $a=2$ \citep{RLR02,ZM02}. In the
following, we allow $b$ to vary and find the joint constraint that
the \citet{Frail01} relation puts on $a$ and $b$.

The afterglow emission becomes prominent when the GRB ejecta
sweeps enough of the external medium to decelerate significantly.
After this time most of the energy is in the shocked external
medium, and energy conservation implies
\begin{equation}
\epsilon \approx \Gamma^2 \mu c^2 = \Gamma^2 Ac^2R^{3-k}/(3-k)\ ,
\end{equation}
where $\mu$ is the swept-up rest mass per solid angle.

In the UJ model $\Gamma(t_j) \sim 1/\theta_j$
\citep{Rhoads97,SPH99} while in the USJ model $\Gamma(t_j) \sim
1/\theta_{\rm obs}$ (Rossi, Lazzati \& Rees 2002; Zhang \&
M\'esz\'aros 2002) and the emission around $t_j$ is dominated by
material close to our line of sight. Together with the relation $R
\sim 4\Gamma^2 ct$, we have
\begin{equation}
t_j
\approx\frac{1}{4c}\left[\frac{(3-k)\epsilon(\theta)}{Ac^2}\right]^{1/(3-k)}
\theta^{2(4-k)/(3-k)}
    \propto \theta^{[2(4-k)-a]/(3-k)}\ ,
\end{equation}
where $\theta=\theta_j$ for the UJ model and $\theta=\theta_{\rm
obs}$ for the USJ model. Now, in order to satisfy the observed
correlation\footnote{We have
$E_{\rm\gamma,iso}^{-1}\propto\theta^{a+b}$ since
$E_{\rm\gamma,iso}(\theta)=4\pi\epsilon(\theta)\epsilon_\gamma(\theta)$.
The peak isotropic equivalent luminosity is given by
$L=E_{\rm\gamma,iso}/T$ where $T$ is the effective duration of the
GRB, which is assumed to be uncorrelated with $L$ so that $L$ has
the same $\theta$-dependence as $E_{\gamma,{\rm iso}}$},
$t_j\propto E_{\rm\gamma,iso}^{-1}\propto\theta^{a+b}$, we must
have
\begin{equation}\label{lambda}
\lambda\equiv a+b = \frac{2(4-k)-a}{3-k}\ .
\end{equation}
Therefore if we allow the efficiency and the energy per solid
angle to vary in this way, we obtain the condition
$b=(2-a)(4-k)/(3-k)$ or $a=2-b(3-k)/(4-k)$ in order to reproduce
the observed correlation of \citet{Frail01}.

For the USJ model, interesting constraints on the power-law index
$a_{\rm USJ}$ of the energy per solid angle have been derived from
the shape of the afterglow light curve \citep{GK03,KG03}. For
$a_{\rm USJ}\lesssim 1.5$ the change in the temporal decay index
across the jet break is too small compared to observations, while
for $a_{\rm USJ}\gtrsim 2.5$ there is a very pronounced flattening
of the light curve before the jet break, which is not observed.
The latter feature arises since the inner parts of the jet near
the core, where $\epsilon$ is the largest, become visible. This
explains why this feature becomes more pronounced for $a_{\rm
USJ}>2$, where most of the energy in the jet is concentrated at
small angles, near the core. There is no sharp borderline where
the light curves change abruptly. Instead, the light curves change
smoothly with the parameter $a_{\rm USJ}$. Altogether, the
observed shape of the afterglow light curve constrains the
parameter $a_{\rm USJ}$ to be in the range $1.5\lesssim a_{\rm
USJ} \lesssim 2.5$. It is important to stress that this constraint
applies only to the USJ model and not to the UJ model, since in
the former $a_{\rm USJ}$ determines the structure of an individual
jet (and therefore affects the light curves) while in the latter
the structure of an individual jet is fixed and $a_{\rm UJ}$ only
affects how the uniform $\epsilon$ changes between different jets
of different half-opening angles $\theta_j$.

One might also try to constrain the power-law index $b$ of the
gamma-ray efficiency $\epsilon_\gamma$ from observations. This
would, of course, apply to both the USJ model and the UJ model.
The way in which this has been done so far is by taking
$\epsilon_\gamma=E_{\gamma,{\rm iso}}/(E_{\gamma,{\rm iso}}+E_{\rm
k,iso})$ where $E_{\rm k,iso}$ is the initial value of the
isotropic equivalent kinetic energy. However, it is hard to
evaluate $E_{\rm k,iso}$ very accurately. It is usually evaluated
from afterglow observations several hours to days after the GRB,
and can provide an estimate accurate to within a factor of $\sim
2$ or so for the kinetic energy at that time. An additional and
potentially larger uncertainty arises since at early times there
is fast cooling and radiative losses might reduce the initial
kinetic energy by up to an order of magnitude or so. This can be
taken into account, but introduces an additional uncertainty in
the value of $E_{\rm k,iso}$ that is estimated in this way.

\citet{PK01,PK02} evaluated $E_{\rm k,iso}$ from a fit to the
broadband afterglow data for ten different GRBs and obtained
$\epsilon_\gamma\gtrsim 0.5$ for most of these GRBs and
$\epsilon_\gamma\gtrsim 0.1$ for all of them. There seems to be no
particular correlation with $\theta$, but due to the small number
of bursts and the reasonably large uncertainty in $E_{\rm k,iso}$
there might still be some intrinsic correlation. \citet{L-RZ04}
used the X-ray luminosity at $10\;$hr to estimate the kinetic
energy at that time and used a simple analytic expression
\citep{Sari97} to account for the radiative losses.  In this way
they estimated $\epsilon_\gamma$ for 17 GRBs and one X-ray flash
(XRF 020903). Their Figure 6 suggests that $\epsilon_\gamma$
decreases with $\theta$, although they claim that this is not
statistically significant. Since most of the estimates for
$\epsilon_\gamma$ are $\gtrsim 0.5$ and are for small values of
$\theta$, and since $\epsilon_\gamma<1$ by definition, this
suggests $b\gtrsim 0$ (otherwise we would have $\epsilon_\gamma>1$
at large $\theta$, which is impossible). Moreover, Figure 6 of
\citet{L-RZ04} suggests that $0\lesssim b\lesssim 1$. If we assume
that $0\lesssim b\lesssim 1$, then Eq. (\ref{lambda}) would imply
$1.25\lesssim a\lesssim 2$, $2\lesssim\lambda\lesssim 2.25$ for
$k=0$, and $1.5\lesssim a\lesssim 2$, $2\lesssim\lambda\lesssim
2.5$ for $k=2$.

\section{Luminosity Function for Different Jet Structures}
\label{lum_fun}

For simplicity, we assume that the emission from the GRB jet is
axially symmetric, so that the observed luminosity depends only on the
viewing angle $\theta$ from the jet axis, and does not depend on the
azimuthal angle $\phi$. It is convenient to define the probability
distribution $P(L|\theta)$, where $P(L|\theta)dL$ is the probability
for an isotropic equivalent luminosity between $L$ and $L+dL$, when
viewing the jet from an angle $\theta$.  Naturally, we must have $\int
P(L|\theta)dL=1$ for any value of $\theta$.  The probability of
viewing the jet from an angle between $\theta$ and $\theta+d\theta$ is
simply $P(\theta)d\theta=\sin\theta d\theta$, and
$\int_0^{\pi/2}P(\theta)d\theta=1$. Averaging over the viewing angle
$\theta$ we obtain the overall probability distribution for $L$,
\begin{equation}\label{P(L)}
P(L)=\int_0^{\pi/2}
P(\theta)P(L|\theta)d\theta=\int_0^1P(L|\theta)\, d\cos\theta\ .
\end{equation}
For a universal GRB jet structure, i.e., if all GRB jets have the
same intrinsic properties, then Eq. (\ref{P(L)}) represents the
GRB luminosity function.

As a simple and very useful example, let us consider a universal
structured jet model where\footnote{This is the standard universal
structured jet model \citep{RLR02,ZM02} for $\lambda=2$, and
assuming no scatter in $L$ for a given $\theta$.}
$P(L|\theta)=\delta[L-L(\theta)]$ and
$L(\theta)=L_0(\theta/\theta_0)^{-\lambda}$. Using Eq.
(\ref{P(L)}) this implies $P(L)=(\theta_0/\lambda
L_0)(L/L_0)^{-1-1/\lambda}\sin\theta(L)$, where
$\theta(L)=\theta_0(L/L_0)^{-1/\lambda}$, and for $\theta(L)\ll 1$
it reduces to $P(L)\approx(\theta_0^2/\lambda
L_0)(L/L_0)^{-1-2/\lambda}$. Alternatively, one may
assume\footnote{This is a pure power law in $L$ for all $\theta$,
instead of just for $\theta\ll 1$ as we had before.} $P(L)=C_0
L^{-\eta}$, where $\eta=1+2/\lambda$ and
$C_0\sim\lambda^{-1}\theta_0^2L_0^{\eta-1}$ is determined by the
normalization condition, $\int P(L)dL=1$.

An important point to stress here is that $P(L)$ represents the
average over all possible viewing angles. Even if the jet is
assumed to have a sharp outer edge at some finite angle
$\theta_{\rm max}$, with some probability distribution $P^*(L)$
for $\theta<\theta_{\rm max}$, such that
$P^*(L)=(1-\cos\theta_{\rm max})^{-1}\int_0^{\theta_{\rm
max}}d\theta\sin\theta P(L|\theta)$ and
$P(L|\theta>\theta_j)=\delta(L)$, then $P(L)=(1-\cos\theta_{\rm
max})P^*(L)+\cos\theta_{\rm max}\delta(L)$.

If the GRB jet structure is not universal, and $P(L|\theta)$
depends on additional parameters that describe the jet structure,
then Eq. (\ref{P(L)}) needs to be averaged over these parameters
in order to obtain the GRB luminosity function. For example, for a
uniform jet of half-opening angle $\theta_j$, we have
$P(L|\theta,\theta_j)=\Theta(\theta_j-\theta)P^*(L|\theta_j)
+\Theta(\theta-\theta_j)\delta(L)$ where $P^*(L|\theta_j)$ is the
probability distribution of $L$ inside the jet (at
$\theta<\theta_j$) for a given value of $\theta_j$. In this case,
if $P(\theta_j)$ is the probability distribution for $\theta_j$,
then the GRB luminosity function is given by
\begin{eqnarray}\nonumber
P(L) &=& \int_0^{\pi/2}P(\theta_j)d\theta_j \int_0^{\pi/2}
P(\theta)d\theta\,P(L|\theta,\theta_j)
\\ \label{P(L)2}
 &=& \int_0^{\pi/2}P(\theta_j)d\theta_j
\left[(1-\cos\theta_j)P^*(L|\theta_j)+\cos\theta_j\delta(L)\right]\
.
\end{eqnarray}
In analogy with the USJ model, we assume for the UJ model that
$\epsilon=\epsilon_0(\theta_j/\theta_{\rm min})^{-a_{\rm UJ}}$ and
$\epsilon_\gamma=\epsilon_{\gamma,0}(\theta_j/\theta_{\rm
min})^{-b_{\rm UJ}}$. Note that since $\epsilon=E_{\rm
iso}/4\pi\approx E/2\pi\theta_j^2$ this implies that the true energy
is not necessarily constant, $E\propto\theta_j^{2-a_{\rm UJ}}$. In
order to imitate a structured jet with a uniform core, we chose
\begin{equation}\label{PLth_UJ}
P(\theta_j)=\Theta(\theta_j-\theta_{\rm min})\Theta(\theta_{\rm
max}-\theta_j)C(q)\theta_j^{-q}+B(q)\delta(\theta_j-\theta_{\rm
min})\ ,
\end{equation}
where $C(q)=[1-B(q)](1-q)/(\theta_{\rm max}^{1-q}-\theta_{\rm
min}^{1-q})$ for $q\neq 1$ and $C(1)=[1-B(1)]/\ln(\theta_{\rm
max}/\theta_{\rm min})$ from the normalization $\int
P(\theta_j)d\theta_j=1$. For
$P^*(L|\theta_j)=\delta[L-L(\theta_j)]$ where $L(\theta_j)=L_{\rm
min}(\theta_j/\theta_{\rm max})^{-\lambda_{\rm UJ}}$ and $L_{\rm
min}=L(\theta_{\rm max})$, this implies
\begin{eqnarray}\nonumber
P(L)&=&\Theta(L-L_{\rm min})\Theta (L_{\rm max}-L)\frac{C\theta_{\rm
max}^{1-q}}{\lambda_{\rm UJ}L_{\rm min}}[1-\cos\theta_j(L)]
\\ \nonumber
& &\times\left(\frac{L}{L_{\rm min}}\right)^{-1-(1-q)/\lambda_{\rm
UJ}} +B(1-\cos\theta_{\rm min})\delta(L-L_{\rm max})
\\ \nonumber
& &+\ \left[B\cos\theta_{\rm min}+C\int_{\theta_{\rm
min}}^{\theta_{\rm max}}
d\theta\,\theta^{-q}\cos\theta\right]\delta(L)
\\ \nonumber
&\approx &\Theta(L-L_{\rm min})\Theta (L_{\rm
max}-L)\frac{C\theta_{\rm max}^{3-q}}{2\lambda_{\rm UJ}L_{\rm
min}}\left(\frac{L}{L_{\rm min}}\right)^{-1-(3-q)/\lambda_{\rm
UJ}}
\\ \nonumber
& & +B\frac{\theta_{\rm min}^2}{2}\delta(L-L_{\rm max})
\\ \label{uniform_P(L)}
& &+\left[1-B\frac{\theta_{\rm min}^2}{2}-\frac{C}{2}
\frac{(\theta_{\rm max}^{3-q}-\theta_{\rm min}^{3-q})}{(3-q)}
\right]\delta(L)\ ,
\end{eqnarray}
\begin{eqnarray}\nonumber
P^*(L)&=& \Theta(L-L_{\rm min})\Theta (L_{\rm
max}-L)\frac{C\theta_{\rm max}^{1-q}}{\lambda_{\rm UJ}L_{\rm
min}}\left(\frac{L}{L_{\rm min}}\right)^{-1-(1-q)/\lambda_{\rm
UJ}}
\\
& &+B\delta(L-L_{\rm max}) \ ,
\end{eqnarray}
where $\theta_j(L)=\theta_{\rm max}(L/L_{\rm
min})^{-1/\lambda_{\rm UJ}}$.

In the case of the  structured jet discussed above, with a sharp
outer edge at some finite angle $\theta_{\rm max}$, and a uniform
core within some angle $\theta_c$, we have
\begin{eqnarray}\nonumber
P(L|\theta)&=& \Theta(\theta_c-\theta)\delta(L-L_{\rm
max})+\Theta(\theta-\theta_{\rm max})\delta(L)
\\ \label{PLth_USJ}
& &+\Theta(\theta-\theta_c)\Theta(\theta_{\rm
max}-\theta)\delta[L-L(\theta)]\ ,
\end{eqnarray}
 where $L(\theta)=L_{\rm min}(\theta/\theta_{\rm
max})^{-\lambda_{\rm USJ}}$ and $L_{\rm min}=L(\theta_{\rm max})$.
Therefore,
\begin{eqnarray}\nonumber
P(L)&=&\Theta(L_{\rm max}-L)\Theta(L-L_{\rm min})\frac{\theta_{\rm
max}\sin\theta(L)}{\lambda_{\rm USJ}L_{\rm
min}}\left(\frac{L}{L_{\rm min}}\right)^{-1-1/\lambda_{\rm USJ}}
\\ \nonumber
& & +\;(1-\cos\theta_c)\delta(L-L_{\rm
max})+\delta(L)\cos\theta_{\rm max}
\\ \label{structure}
&\approx & \Theta(L_{\rm max}-L)\Theta(L-L_{\rm
min})\frac{\theta_{\rm max}^2}{\lambda_{\rm USJ}L_{\rm
min}}\left(\frac{L}{L_{\rm min}}\right)^{-1-2/\lambda_{\rm USJ}}
\\ \nonumber
& & +\;\frac{\theta_c^2}{2}\delta(L-L_{\rm
max})+\left(1-\frac{\theta_{\rm max}^2}{2}\right)\delta(L)\ ,
\\ \nonumber
 P^*(L)&=&\frac{\Theta(L_{\rm max}-L)\Theta(L-L_{\rm min})}{(1-\cos\theta_{\rm
max})\theta_{\rm max}^{-1}\lambda_{\rm USJ}L_{\rm
min}}\sin\theta(L)\left(\frac{L}{L_{\rm
min}}\right)^{-1-1/\lambda_{\rm USJ}}
\\ \nonumber
& &+\frac{(1-\cos\theta_c)}{(1-\cos\theta_{\rm max})}\delta(L-L_{\rm
max})
\\ \nonumber
&\approx &\Theta(L_{\rm max}-L)\Theta(L-L_{\rm
min})\frac{2}{\lambda_{\rm USJ}L_{\rm min}}\left(\frac{L}{L_{\rm
min}}\right)^{-1-2/\lambda_{\rm USJ}}
\\
& &+\left(\frac{\theta_c}{\theta_{\rm max}}\right)^2\delta(L-L_{\rm
max})\ ,
\end{eqnarray}
where $\theta(L)=\theta_{\rm max}(L/L_{\rm min})^{-1/\lambda_{\rm
USJ}}$ and $\int_{L_{\rm min}}^\infty P^*(L)dL=1$ as it should be.
Note that the coefficient in the first term in the expression for
$P(L)$ includes $\theta_{\rm max}$ only through the combination
$\theta_{\rm
  max}L_{\rm min}^{1/\lambda}=\theta_{\rm max}[L(\theta_{\rm
  max})]^{1/\lambda}$, which is independent of $\theta_{\rm max}$ (for
a fixed normalization of $L$) since
$L(\theta)\propto\theta^{-\lambda}$.

By comparing equations (\ref{uniform_P(L)}) and (\ref{structure}) one
can also see that a similar luminosity function can be obtained
for different values of $\lambda=a+b$, as long as $\lambda_{\rm
UJ}/\lambda_{\rm USJ}=(3-q)/2$. In this case, when $q\neq 1$ then
$\lambda_{\rm UJ}\neq\lambda_{\rm USJ}$. For the same $\theta_{\rm
max}$, $L_{\rm min}$ and $L_{\rm max}$ we have $L_{\rm max}/L_{\rm
min}=(\theta_{\rm max}/\theta_{\rm min})^{\lambda_{\rm
UJ}}=(\theta_{\rm max}/\theta_c)^{\lambda_{\rm USJ}}$ which
implies $\theta_c/\theta_{\rm min}=(\theta_{\rm min}/\theta_{\rm
max})^{(1-q)/2}$ (where $\theta_c<\theta_{\rm min}$ for $q<1$ and
$\theta_c>\theta_{\rm min}$ for $q>1$). In order for the
luminosity functions to be the same also for the core of the
structured jet, which is represented by the term
$\propto\delta(L-L_{\rm max})$, the ratio of this term to the
other terms should be the same for equations (\ref{uniform_P(L)}) and
(\ref{structure}). Thus we obtain
\begin{eqnarray}\label{A}
C(q) &=&\left[\frac{\theta_{\rm
min}^{1-q}}{(3-q)}+\frac{\theta_{\rm max}^{1-q}-\theta_{\rm
min}^{1-q}}{(1-q)}\right]^{-1}\ ,
\\ \label{B}
B(q) &=&\left\{1+\frac{(3-q)}{(1-q)}\left[\left(\frac{\theta_{\rm
max}}{\theta_{\rm min}}\right)^{1-q}-1\right]\right\}^{-1}\ ,
\end{eqnarray}
where $B(1)=C(1)/2=[1+2\ln(\theta_{\rm max}/\theta_{\rm
min})]^{-1}$ and $C(q)/B(q)= (3-q)\theta_{\rm min}^{q-1}$.

\begin{figure}[b]
\plotone{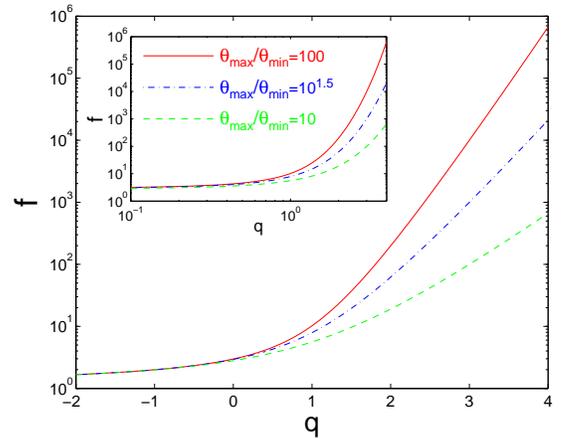} \caption{\label{fig_f(q)} The ratio $f$ (Eq.
  \ref{f}) of the true GRB rate for the UJ model to the true GRB rate
  for the USJ model, given the same observed GRB rate and luminosity
  function (i.e., the same $\log N-\log S$ distribution), as a
  function of the slope $q$ of the probability distribution for the
  opening angle $\theta_j$ of the uniform jet,
  $P(\theta_j)\propto\theta_j^{-q}$ for $\theta_{\rm
    min}<\theta_j<\theta_{\rm max}$. See text for details.}
\end{figure}

The normalization of the first two terms in Eq.
(\ref{uniform_P(L)}) for the UJ model is smaller than that of the
(corresponding) first two terms in Eq. (\ref{structure}) for the
USJ model by a factor of
\begin{eqnarray}\nonumber
f(q)&=&\left(\frac{\theta_{\rm min}}{\theta_{\rm
max}}\right)^{1-q}+\frac{(3-q)}{(1-q)}\left[1-\left(\frac{\theta_{\rm
min}}{\theta_{\rm max}}\right)^{1-q}\right]
\\ \label{f}
&\approx &\left\{\begin{matrix} (3-q)/(1-q) & \ \ (q<1)\ , \cr & \cr
[2/(q-1)](\theta_{\rm max}/\theta_{\rm min})^{q-1} & \ \ (q>1)\ ,
\end{matrix}\right.
\end{eqnarray}
where $f(1)=1+2\ln(\theta_{\rm max}/\theta_{\rm min})$. This
factor $f$ is the ratio of the true GRB rate for the UJ model to
the true GRB rate for the USJ model that corresponds to the same
luminosity function (i.e. Eqs. \ref{uniform_P(L)} and
\ref{structure}). Figure \ref{fig_f(q)} shows $f(q)$ for three
different values of $\theta_{\rm max}/\theta_{\rm min}$. The fact
that $f(q)>1$ means that a larger number of uniform jets is
needed, compared to structured jets, in order to reproduce the
same luminosity function and the same observed rate of GRBs,
$\dot{N}_{\rm GRB}$. This means that for the same $\dot{N}_{\rm
GRB}$ the intrinsic GRB rate per unit comoving volume, $R_{\rm
GRB}(z)$, and specifically $R_{\rm GRB}(z=0)$, is larger by a
factor of $f$ for the UJ model compared to the USJ model.

The same factor $f$ is obtained from the ratio of the energy
output in gamma-rays of a single structured jet to the average
energy output in gamma-rays of a uniform jet. This ratio must
equal the ratio of the intrinsic rates since the total energy
output in gamma-rays per unit time per unit volume must be the
same \citep{GPW04}.

We can see that $f\sim 1$ for $q<1$ and $f\sim(\theta_{\rm
max}/\theta_{\rm min})^{q-1}$ for $q>1$. This can be understood as
follows. Since most of the solid angle of the structured jet is
near $\theta_{\rm max}$, the number of uniform jets with
$\theta_j\sim\theta_{\rm max}$ should be comparable to the number
of structured jets. Therefore, for $q<1$ where most of the uniform
jets have $\theta_j\sim\theta_{\rm max}$ we have $f\sim 1$, while
for $q>1$ where most of the uniform jets have
$\theta_j\sim\theta_{\rm min}$, $f\sim(\theta_{\rm
max}/\theta_{\rm min})^{q-1}$ is roughly the inverse of the
fraction of jets with $\theta_j\sim\theta_{\rm max}$.

We have demonstrated that the luminosity function of the USJ model
can be imitated by the UJ model with the appropriate choice of
$P(\theta_j)$. The implied true GRB rate $R_{\rm GRB}(z)$ for the
same observed GRB rate $\dot{N}_{\rm GRB}$ would always be larger
(by a factor $f$) for the UJ model. For the UJ model, the term
$B\delta(\theta_j-\theta_{\rm min})$ in $P(\theta_j)$ which
produces the term $\propto\delta(L-L_{\rm max})$ in $P(L)$ is
somewhat artificial, and was introduced only to make a complete
analogy with $P(L)$ for the USJ model. The exact form of the
cutoff near $\theta_{\rm min}$ is not clear and probably would not
have a very large effect on the $\log N - \log S$ distribution,
making it hard to distinguish between the two models in this way.
It is also important to keep in mind that the observational
constraints on the parameter $a$ from the shape of the afterglow
light curve, namely $1.5\lesssim a_{\rm USJ} \lesssim 2.5$, apply
only to the USJ model and not to the UJ model.

The form of the luminosity function given in equations
(\ref{uniform_P(L)}) or (\ref{structure}) is valid only if
$\theta^{\lambda}L(\theta)\sim{\rm const}$, where
$\theta=\theta_{\rm obs}$ for the USJ model and $\theta=\theta_j$
for the UJ model. How well this condition is satisfied may be
tested by using the values of both $L$ and $\theta$ that were
estimated from observations for a sample of GRBs. We consider a
sub-sample of 19 GRBs out of the \citet{BFK03} sample, for which
there is both a known redshift and an estimate for $\theta$. There
are larger uncertainties in the determination of the peak
luminosity,\footnote{We use the peak luminosity since it is
usually the most relevant quantity for the triggering of the
different detectors.} as the bursts were detected by different
instruments with different temporal and spectral sensitivities.
Following \citet{GPW04} we have extrapolated the GRB fluxes to the
BATSE range ($50\;{\rm keV}-300\;$keV) using the method described
in \citet{SB01}. In our analysis we have used the median value of
the spectral photon index in the $50\;{\rm keV}-300\;$keV band for
the long bursts sample, $-1.6$, as found by \citet{Schmidt01}. The
redshifts and fluxes were taken from the table given in
\citet{VPR03}.

We calculate the values of $\theta^\lambda L$ for such a sample
and for different values of $\lambda$ within the range allowed for
the USJ model from the constraints given above (namely,
$1.5\lesssim a_{\rm USJ}\lesssim 2.5$ and Eq. \ref{lambda}):
$1.83\lesssim\lambda\lesssim 2.17$ for $k=0$,  and
$1.5\lesssim\lambda\lesssim 2.5$ for $k=2$. Therefore overall
$1.5<\lambda<2.5$. The result of this analysis is that the
distribution of $\theta^\lambda L(\theta)$ is not quite a delta
function, and there is some dispersion around the mean value. This
dispersion is reasonably fit by a log-normal distribution,
\begin{equation}
\label{P(L,theta)}
P(L|\theta)=\frac{1}{L\sigma\sqrt{2\pi}}\times\exp\left\{-\frac{\left[\ln
L - \ln L(\theta)\right]^2}{2\sigma^2}\right\}\ ,
\end{equation}
with $L(\theta) = L_0(\theta/\theta_0)^{-\lambda}$, where
$\theta_0^{\lambda}L_0$ is the average value of
$\theta^{\lambda}L(\theta)$ for the observed sample, and $\sigma$
is the standard deviation of $\ln[\theta^{\lambda}L(\theta)]$,
which is determined by a fit to the dispersion of the observed
sample. This distribution approaches a delta function (i.e. with
no scatter) for $\sigma\to 0$.

We performed fits of $\theta^\lambda L$ to a sample of 19 bursts
from \citet{BFK03} which were observed by BATSE and Beppo-SAX. For
$\lambda=(1.5,\,2,\,2.5,\,3)$ we obtained
$\theta_0^{\lambda}L_0=(15,\,5.8,\,2.2,\,1.1)\times 10^{49}\;{\rm
erg\;s^{-1}}$ and $\sigma_{\rm obs}=(1.2,\,1.2,\,1.3,\,1.5)$. These
values of $\sigma_{\rm obs}$ can be thought of as upper limits on
$\sigma$ (i.e. $\sigma\lesssim\sigma_{\rm obs}$), since the
observed scatter in $\theta^{\lambda}L(\theta)$ includes both the
intrinsic scatter (which should be reflected in $\sigma$) and
additional scatter due to measurement errors in $\theta$ (which
produces some scatter $\sigma_{\rm err}$), which can be as high as
tens of percent. \citet{BFK03} obtain a factor of $2.2$ for the
scatter in $\epsilon\,\theta^2$, implying that
$\sigma_{\ln\theta}\lesssim 0.4$, and therefore $\sigma_{\rm
err}\lesssim 0.4\lambda$. \footnote{This is since the dominant
measurement error in $\theta^\lambda L$ comes from
$\theta^\lambda$, while $L$ is measured more accurately.} Assuming
that the measurement errors in $\theta$ are uncorrelated with the
intrinsic scatter in $\theta^{\lambda}L(\theta)$, this implies
$\sigma_{\rm obs}^2\approx\sigma^2+\sigma_{\rm
err}^2\lesssim\sigma^2+(0.4\lambda)^2$and therefore
$(1.0,\,0.9,\,0.8,\,0.9)\lesssim\sigma\lesssim(1.2,\,1.2,\,1.3,\,1.5)$ for
$\lambda=(1.5,\,2,\,2.5,\,3)$.

The luminosity function in this case can be obtained using equations
(\ref{P(L)}) and (\ref{P(L,theta)}). It is important to note that
the integral in Eq. (\ref{P(L)}) must be done over all possible
viewing angles, even if the jet is assumed to have a core angle,
$\theta_c$, and maximal angle, $\theta_{\rm max}$. In fact,
together with the power-law index ($\lambda$) and normalization
($\theta_0^\lambda L_0$), the other two intrinsic parameters that
define a power-law universal structured jet are the angles of its
outer edge ($\theta_{\rm max}$) and inner core ($\theta_c$). For
the UJ model outlined above, $\theta_{\rm max}$ and $\theta_c$ are
replaced by $\theta_{\rm max}$ and $\theta_{\rm min}$ which
determine the range of possible $\theta_j$ values, where
$P(\theta_j)\propto\theta_j^{-q}$ also depends on an additional
parameter $q$.

The observed rate of GRBs (over the entire sky) with a peak photon
flux greater than $S$ is given by:
\begin{equation}\label{iso}
 \dot{N}_{\rm GRB}(>S)=\int P(L)dL\int_0^{z_{\rm max}(L,S)} \frac{R_{\rm
GRB}(z)}{(1+z)}\frac{dV(z)}{dz}dz\ ,
\end{equation}
where $P(L)$ is given by equations (\ref{P(L)}) and (\ref{P(L,theta)}),
$z$ is the redshift of the GRB, $z_{\rm max}(L,S)$ is the maximal
redshift from which a GRB with luminosity $L$ and peak flux $S$
can be detected, and $R_{\rm GRB}(z)$ is the (true) GRB rate per
unit comoving volume $V$.\footnote{Note that $R_{\rm GRB}(z)$ is
the true GRB rate, and it is obtained directly from this formalism
without any need for a `correction factor' etc.} The factor
$(1+z)^{-1}$ accounts for the cosmological time dilation and
$dV(z)/dz$ is the comoving volume element.

\section{Constraints on the luminosity function of the USJ model}
\label{USJ}

We consider all the GRBs in the GUSBAD catalog \citep{Schmidt04},
which lists 2204 GRBs detected at a time scale of $1024\;$ms. This
catalog contains all the long GRBs \citep[$T_{90}>
2\;$sec;][]{K93}, detected while the BATSE on-board trigger
\citep{P99} was set for 5.5 $\sigma$ over background in at least
two detectors, in the energy range $50-300\;$keV. Using this
sample we estimate the ratio $C_{\rm max}/C_{\rm min}$ for each
burst, where $C_{\rm max}$ is the count rate in the second
brightest illuminated detector and $C_{\rm min}$ is the minimum
detectable rate. We find $\langle V/V_{\rm max} \rangle=0.335\pm
0.007$.

We consider also the Rowan-Robinson SFR \citep[RR-SFR;][]{RR99}
that can be fitted with the expression
\begin{equation}
\label{RR} R_{\rm GRB}(z) = \rho_0\times\left\{ \begin{array}{ll}
10^{0.75 z} & z<1 \ ,\\
10^{0.75 } & z>1\ ,
\end{array}
\right.
\end{equation}
where $\rho_0=R_{\rm GRB}(z=0)$ is the true GRB rate per unit comoving
volume at z=0. Throughout the paper we use the following cosmological
parameters: ($\Omega_{\rm M},\Omega_\Lambda,h)=(0.3,0.7,0.7)$.

Objects with luminosity $L$ observed by BATSE with a flux limit
$S_{\rm lim}$ are detectable to a maximum redshift $z_{\rm
  max}(L,S_{\rm lim})$. The limiting flux has a distribution $G(S_{\rm
  lim}$) that can be obtained from the distribution of $C_{\rm min}$
of the GUSBAD catalog. Considering five main representative intervals
we obtain that $30\%$, $30\%$, $10\%$, $20\%$, and $10\%$ of the
sample have $S_{\rm lim}\sim 0.23$, $0.25$, $0.22$, $0.26$, and
$0.27\;{\rm ph\;cm^{-2}\;s^{-1}}$, respectively.  Therefore, we have
\begin{eqnarray}
\nonumber \dot{N}_{\rm
GRB}(>S)=\quad\quad\quad\quad\quad\quad\quad\quad\quad\quad
\\ \nonumber
\int P(L)dL \int G(S_{\rm lim}) dS_{\rm lim} \int_0^{z_{\rm
max}[L,\max(S,S_{\rm lim})]} \frac{R_{\rm GRB}(z)}{(1+z)}
\frac{dV(z)}{dz}dz
\\ \nonumber
=\int P(L)dL \left[\int_0^{S} G(S_{\rm lim}) dS_{\rm lim}
\int_0^{z_{\rm max}(L,S)} \frac{R_{\rm
GRB}(z)}{(1+z)}\frac{dV(z)}{dz}dz \right.\quad
\\
+ \left. \int_{S}^{\infty} G(S_{\rm lim}) dS_{\rm lim}
\int_0^{z_{\rm max}(L,S_{\rm lim})} \frac{R_{\rm GRB}(z)}{(1+z)}
\frac{dV(z)}{dz}dz\right]\ ,\quad\ \;
\end{eqnarray}
where $P(L)$ is given by equation (\ref{P(L)}). For the USJ model
$P(L|\theta>\theta_{\rm max})=\delta(L)$ while
$P(L|\theta<\theta_{\rm max})$ is given by (\ref{P(L,theta)}) with
$L(\theta)=L_{\rm max}\times\min[1,(\theta/\theta_{\rm
min})^{-\lambda_{\rm USJ}}]$ where $L_{\rm max}=L_{\rm
min}(\theta_{\rm min}/\theta_{\rm max})^{-\lambda_{\rm USJ}}$. A
similar dispersion is introduced in the UJ model. The predicted
$\log N-\log S$ distribution depends on the values of $\lambda_{\rm
USJ}$, $\theta_c$, $\theta_{\rm max}$, $\sigma$ and the
normalization $\theta_0^{\lambda_{\rm USJ}} L_0$, where the last
parameter is determined through a fit to observations and is not
considered to be a free parameter. The scatter $\sigma$ is
constrained by observations,
$(1.0,\,0.9,\,0.8,\,0.9)\lesssim\sigma\lesssim(1.2,\,1.2,\,1.3,\,1.5)$
for $\lambda_{\rm USJ}=(1.5,\,2,\,2.5,\,3)$, but was allowed to vary
over a wider range when performing the fit to the data. The smallest
observed values of $\theta$ provide an upper limit on $\theta_c$ of
$\theta_c\lesssim 0.05$ while the largest observed values of
$\theta$ provide a lower limit on $\theta_{\rm max}$ of $\theta_{\rm
max}\gtrsim 0.5$. For the USJ model the shape of the afterglow light
curves implies $1.5\lesssim a_{\rm USJ}\lesssim 2.5$ and therefore
$1.5\lesssim\lambda_{\rm USJ}\lesssim 2.5$, however, since this
limit does not apply to the UJ model and since we wanted to properly
check the consistency of the USJ with the data we allowed
$\lambda_{\rm USJ}$ to vary over a wider range when performing the
fit to the data.

\newcommand{\rb}[1]{\raisebox{1.5ex}[0pt]{#1}}
\newcommand{\rbp}[1]{\raisebox{0.5ex}[0pt]{#1}}
\newcommand{\rbm}[1]{\raisebox{-0.5ex}[0pt]{#1}}
\begin{deluxetable}{llcc}
\tabletypesize{\footnotesize} \tablecaption{Fits to the $\log N -
\log S$ Distribution} \tablewidth{0pt} \tablehead{\colhead{$P(L)$} &
\colhead{model} & \colhead{$R_{\rm
GRB}(z=0)$} & goodness \\
\colhead{model} & \colhead{parameters} &
\colhead{[Gpc$^{-3}\;$yr$^{-1}$]} & \colhead{of fit}} \startdata
 & $\lambda_{\rm USJ}=2.9_{-0.5}^{+0.2}$ &  & $\chi^2=11.05$
\\
\rb{single} & $\sigma=0.5_{-0.5}^{+0.7}$ &
\rb{$0.86_{-0.05}^{+0.14}$ (USJ)} & (9 d.o.f)
\\
\rb{power} & $\theta_{c}=0.05_{-0.01}^{+0.005}$ &  & $P=0.723$
\\
\rb{law} & $\theta_{\rm max}=0.7\pm 0.2$ & \rb{$5.4_{-1.0}^{+1.8}$
(UJ, $q=1$)} & $(1.10\;\sigma)$
\\ \hline
single  & $\lambda_{\rm USJ}=2$ (fixed) &  & $\chi^2=14.67$
\\
P.L.$\;+$ & $\sigma=0.8_{-0.2}^{+0.4}$ & \rb{$0.75_{-0.06}^{+0.07}$
(USJ)} & (10 d.o.f)
\\
 fixed & $\theta_c=0.03_{-0.01}^{+0.005}$ &  &  $P=0.855$   \\
$\lambda_{\rm USJ}$ & $\theta_{\rm max}=0.5\pm 0.05$  &
\rb{$5.0_{-0.8}^{+1.3}$ (UJ, $q=1$)} &
 ($1.46\;\sigma$)
 \\ \hline
double & \rbm{$\alpha=0.6\pm 0.1$} &  & $\chi^2=10.57$ \\
power &  & $10.7\pm 2.3$ & (10 d.o.f) \\
law + & \rb{$\beta=0.8_{-0.1}^{+0.2}$} & (UJ, $q=1$) & $P=0.608$ \\
$\sigma=0$ & \rbp{$L_{51}=1.6_{-0.6}^{+0.8\;\dagger}$} &
 & $(0.86\;\sigma)$ \\
\enddata
\tablecomments{\label{table1}The best fit parameters with their
$1\;\sigma$ confidence intervals are shown together with the implied
true GRB rate and the goodness of fit. The confidence intervals are
the projection onto the relevant parameter axis of the region in the
multi-dimensional parameter space around the global minimum
$\chi^2_{\rm min}$ of $\chi^2$ where
$\Delta\chi^2=\chi^2-\chi^2_{\rm min}<1$.
\newline
\indent$\ \ \ \,^{\dagger}$Here $L_{51}=L_*/(10^{51}\;{\rm erg\;
s^{-1}})$.}
\end{deluxetable}

\begin{figure}[b]
\plotone{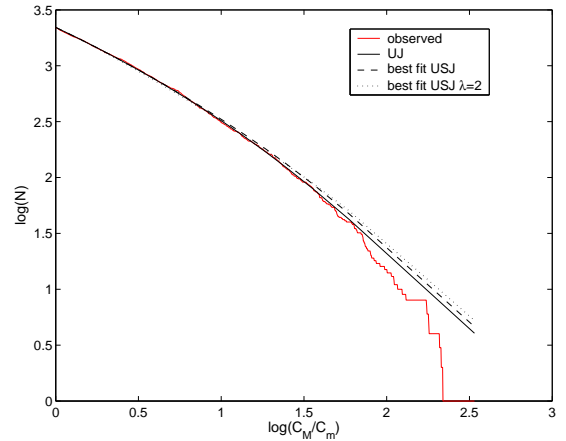} \caption{\label{cumulative} The observed cumulative
  $\log_{10}N-\log_{10}({\rm C}_{\rm max}/{\rm C}_{\rm min})$
  distribution taken from the GUSBAD catalog ({\it solid step-like
    line}) compared to the predicted $\log_{10}N-\log_{10}({\rm
    P}/{\rm P}_{\rm lim})$ distributions for our best fit models: a
  single power-law luminosity function (relevant for both the USJ and
  the UJ models), both when letting all parameters vary ({\it dashes
    line}) and when fixing $\lambda_{\rm USJ}=2$ ({\it dotted line}), and a
  broken power-law luminosity function ({\it solid line}; relevant
  only for the UJ model).}
\end{figure}

Combining the constraint from the Frail relation (Eq.
\ref{lambda}) with the limit on $\lambda$ from the luminosity
function allows us separately to constrain $a$ and $b$. From Eq.
(\ref{lambda}) we have $a = 2(4-k) - (3-k)\lambda = 8 - 3\lambda$
for $k=0$ and $4-\lambda$ for $k=2$. Since for the USJ model $1.5
\la a_{\rm USJ} \la 2.5$ from the afterglow lightcurves, this
implies $11/6 \lesssim \lambda_{\rm USJ} \lesssim 13/6$ for $k=0$
and $1.5 \lesssim \lambda_{\rm USJ} \lesssim 2.5$ for $k=2$. For
$k=0$, $\lambda$ must be close to 2 and $b= (4-k)(\lambda - 2)$
falls in the range $-2/3 \lesssim b\lesssim 2/3$. for $k=2$ this
implies $-1\lesssim b\lesssim 1$. As discussed in \S
\ref{efficiency}, direct estimates of $\epsilon_\gamma$ suggest
$0\lesssim b\lesssim 1$ (this applies both to the USJ model and to the
UJ model). The different constraints on the parameters $a$ and $b$ are
summarized in Fig. \ref{ab}.

If the luminosity function forces $\lambda_{\rm USJ}$ to values
less than 2, then $b$ must be negative and if $\lambda_{\rm USJ}$
approaches 1.5 this would favor an external medium with a $k=2$
density profile. Note that $b <0$ corresponds to the dissipation
efficiency increasing with angle, as would be expected if it is
associated with a shear layer well outside the jet core. We note
that highly relativistic jets passing through an external pressure
gradient less steep than $r^{-4}$ can develop a shocked layer near
the jet boundary, due to the loss of causal contact between the
jet interior and wall.  If this shock is responsible for the
gamma-ray emission, it could lead to radiative efficiency
increasing with $\theta$.

On the other hand, if the luminosity function implies
$\lambda_{\rm USJ}>2$ this would imply $b>0$, i.e. a gamma-ray
efficiency $\epsilon_\gamma$ that decreases with $\theta$, which
is more consistent with direct estimates of $\epsilon_\gamma$. If
$\lambda_{\rm USJ}$ approaches 2.5 this would favor a $k=2$
density profile.

We performed two fits to the data using a single power-law
luminosity function which can arise either in the USJ model or in
the UJ model. In the first fit we let all four free parameters vary:
$\lambda_{\rm USJ}$, $\theta_c$, $\theta_{\rm max}$ and $\sigma$. In
the second fit we held the value of $\lambda$ fixed at $\lambda_{\rm
USJ}=2$ and allowed the remaining three parameters to vary:
$\theta_c$, $\theta_{\rm max}$ and $\sigma$. The second fit was
performed since a value of $\lambda_{\rm USJ}=2$ is expected in the
simplest version of the USJ model where the gamma-ray efficiency
$\epsilon_\gamma$ is constant ($b_{\rm USJ}=0$) and therefore
$a_{\rm USJ}=\lambda_{\rm USJ}=2$ because of the Frail relation (Eq.
\ref{lambda}). Therefore, it is interesting to test whether this
simplest version of the the USJ model is consistent with the
observed $\log N-\log S$ distribution.

In order to assign a $\chi^2$ value to the fit we divided the 2204
GRBs in the GUSBAD catalog into 14 bins according to their value
of $S$, the peak photon flux. The first 11 bins are equally spaced
in $\log S$. In the remaining 3 bins, which correspond to the
highest values of $S$, we chose a larger range of $S$ values so as
to have at least $N_{\rm min}=40$ GRBs in each bin, in order to
have reasonable Poisson statistics. Since the overall
normalization is an additional free parameter in our fits, the
number of degrees of freedom (d.o.f) in our fits is
$\nu=14-(4+1)=9$ in our first fit, and $\nu=10$ in our second fit.

\begin{figure}
\plotone{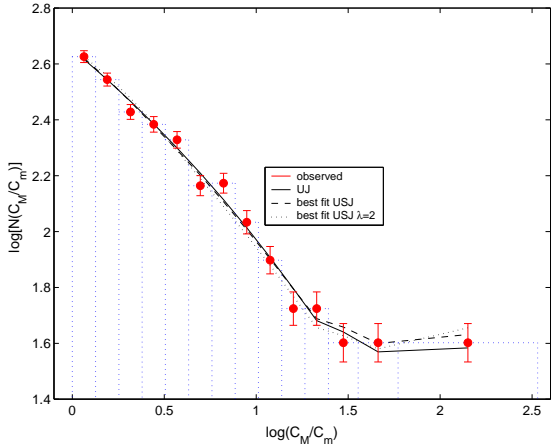} \caption{\label{differential} The same as Figure
\ref{cumulative} but for the
  differential distribution: $\log_{10}({\rm P}/{\rm P}_{\rm lim})$ is
  divided into 14 bins, the first 11 of equal size, and the remaining
  3 with varying sizes chosen such that the number of busts per bin in
  the observed sample is at least $N_{\rm min}=40$ (in order to have
  reasonable statistics so that the Poisson error will not to be too
  large), and $N$ is the observed ({\it circles}) or theoretical ({\it
    lines}; only the values at the center of each bin count) number of
  bursts in each bin. The edges of the bins are also plotted.}
\end{figure}

The results of the fits are presented in Table \ref{table1} and in
Figures \ref{cumulative} and \ref{differential}. When $\lambda$ is
free to vary we obtain a best fit value of $\lambda_{\rm
USJ}=2.9^{+0.2}_{-0.5}$, however, when we fix $\lambda_{\rm
USJ}=2$ we still get an acceptable fit. This suggests that
although values of $\lambda_{\rm USJ}>2$ are preferred by the
data, a value of $\lambda_{\rm USJ}=2$ is still possible. On the
other hand, values of $\lambda_{\rm USJ}<2$ become increasingly
hard to reconcile with the data. Values of $\lambda_{\rm USJ}>2$,
which are preferred by the data, correspond to $b>0$ (a gamma-ray
efficiency $\epsilon_\gamma$ that decreases with $\theta$) and for
the USJ model $\lambda_{\rm USJ}\approx 2.5$ favors $k=2$ (i.e., a
stellar wind external density profile).

The first fit gives $\sigma=0.5_{-0.5}^{+0.7}$ which indicates
that this fit is not very sensitive to the value of $\sigma$. It
also includes the range $0.8\lesssim\sigma\lesssim 1.3$ for
$\lambda_{\rm USJ}=2.5$  and $0.9 \lesssim \sigma \lesssim 1.5$
for $\lambda_{\rm USJ}=2.9$
that are implied by observations. In contrast, the second fit with a
fixed $\lambda_{\rm USJ}=2$ gives $\sigma=0.8^{+0.4}_{-0.2}$ which
requires a positive value of $\sigma$ in order to get a reasonable
fit to the data.

The  best fit values of $\theta_c$ and $\theta_{\rm max}$ are
slightly higher for the first fit, but are rather close between
the two fits. The best fit value of
$\theta_c=0.05_{-0.01}^{+0.005}$ for a free $\lambda$ and
$\theta_c=0.03^{+0.005}_{-0.01}$ for a fixed $\lambda_{\rm USJ}=2$
are close to the upper limit of $\theta_c\lesssim 0.05$ from
observations, and the allowed confidence regions do not allow
values much smaller (by more than a factor of $\sim 2$) than this
limit. The best fit values of $\theta_{\rm max}=0.7\pm 0.2$ for a
free $\lambda$ and $\theta_{\rm max}=0.5\pm0.05$ for a fixed
$\lambda_{\rm USJ}=2$ are close to the lower limit of $\theta_{\rm
max}\gtrsim 0.5$ from observations and are not consistent with
$\theta_{\rm max}=\pi/2$.

The true GRB rate that is implied from our fits is rather close to
the value of $R_{\rm GRB}(z=0)\sim 0.5\;{\rm Gpc^{-3}\;yr^{-1}}$
that was found by \citet{PSF03}. We obtain a slightly larger rate
for the free $\lambda$ fit compared to the fixed $\lambda_{\rm
USJ}=2$ fit, but the difference is very small. The rates we obtain
for the USJ model are lower by a factor of $\sim 300$ compared to
the estimate of \citet{Frail01} for the UJ model.

\begin{figure}
\plotone{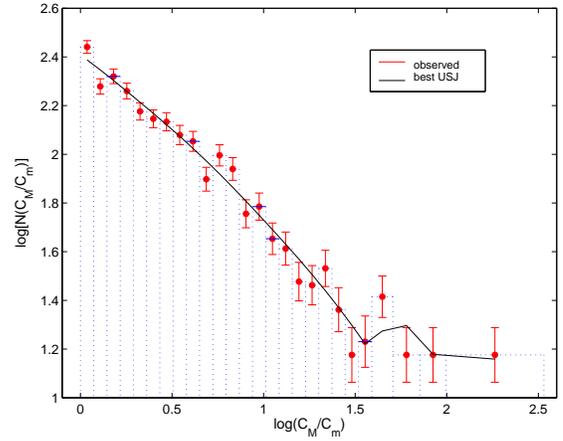} \caption{\label{26bins} The same as Figure
\ref{differential} but for a different binning of $\log_{10}({\rm
P}/{\rm P}_{\rm lim})$: 26 bins and $N_{\rm min}=15$.}
\end{figure}

Finally, since the exact choice for the number and sizes of the bins
used in our fit is somewhat arbitrary and might have some effect on
our results, we estimate this effect more quantitatively by repeating
our fit for a single power law luminosity function with a larger
number of bins, 26 instead of 14 (see Fig. \ref{26bins}). The best fit
parameter values remained the same as for the original binning (14
bins) up to the significant digits shown in Table \ref{table1}, while
the $1\;\sigma$ confidence intervals slightly changed: $\lambda_{\rm
USJ}=2.9_{-0.4}^{+0.2}$, $\sigma=0.5^{+0.7}_{-0.4}$, $\theta_c=0.05\pm
0.05$ and $\theta_{\rm max}=0.7^{+0.2}_{-0.1}$.  The GRB rate is
slightly higher, $R_{\rm GRB}(z=0)=0.95^{+0.25}_{-0.22}\;{\rm
Gpc^{-3}\; yr^{-1}}$, but well within the $1\;\sigma$ confidence
interval for the original binning.  This demonstrates that while the
exact choice of binning has some effect on our results, this effect is
rather small.

In  order  to estimate  the effect  of  our choice  for  the star
formation rate (or more  accurately the GRB   rate that is  assumed
to follow the star formation rate up to an unknown normalization
constant) we repeated our fit  with the  original  binning (14 bins)
but  with a different star formation rate: SFR2 from \citet{PM01}.
We obtained: $\lambda_{\rm USJ}=2.9_{-0.4}^{+0.3}$, $\sigma=0.5\pm
0.5$, $\theta_c=0.04\pm 0.005$ and $\theta_{\rm max}=0.9\pm 0.2$.
While changing the star formation rate has a larger effect than
changing the binning, the parameter values for the two different
star formation rates that we considered are still consistent with
each other within their $1\;\sigma$ confidence intervals. For the
new star formation rate we obtain a slightly lower true GRB rate:
$R_{\rm GRB}(z=0)=0.65_{-0.17}^{+0.05}\; {\rm Gpc^{-3}\;yr^{-1}}$.

\section{The luminosity function for the Uniform Jet Model}
\label{UJ}

The two fits that were discussed in \S \ref{USJ} for the USJ model
still apply to the UJ model with $\theta_c\to\theta_{\rm min}$ and
$\lambda_{\rm USJ}\to\lambda_{\rm UJ}\times 2/(3-q)$. The second
substitution indicates a degeneracy where the luminosity function
$P(L)$ and therefore the $\log N - \log S$ distribution would be
the same for UJ models with the same value of $\lambda_{\rm
UJ}/(3-q)=(a_{\rm UJ}+b_{\rm UJ})/(3-q)$. This degeneracy does not
enable us to determine $P(\theta_j)$, which is parameterized by
$q$ from a fit to the observed $\log N - \log S$ distribution.
Such a fit can only determine $\eta=1+2/\lambda_{\rm
USJ}=1+(3-q)/\lambda_{\rm UJ}$ where $P(L)\propto L^{-\eta}$. The
degeneracy might be broken if we could estimate the true GRB rate
in an independent way. This is since the true GRB rate for the USJ
model is determined from the fit to the observed $\log N - \log S$
distribution, and the correction factor for the UJ model depends
only on $q$ (see Eq. \ref{f} and Figure \ref{fig_f(q)}). Thus an
independent estimate of the true GRB rate would enable the
determination of $q$ and would therefore constrain $P(\theta_j)$.

The value of $\lambda=a+b$ is the same for the UJ and USJ models
if and only if $q=1$. In fact, the Frail relation as manifested in
Eq. (\ref{lambda}) implies that for $q=1$ and a given value of
$k$, it is enough that either $a$ or $b$ be the same for the two
models in order for $\lambda=a+b$ to be the same. Add to this the
fact that for $q=1$ there is an equal number of uniform jets per
logarithmic interval in $\theta_j$, and $q=1$ might be considered
as the most `natural' value for $q$. However, we emphasize that
there is no physical basis for comparing pairs of models with
$\lambda_{\rm USJ}=\lambda_{\rm UJ}$, and therefore no a prior
reason for choosing $q=1$. The only meaningful comparison is to
observations.

The discussion in \S \ref{USJ} about the values of $\lambda_{\rm
USJ}$, $\sigma$, $\theta_{\rm max}$ and $\theta_c$ for the USJ
model are still valid for the UJ model, where $\theta_c$ is
replaced by $\theta_{\rm min}$ and $\lambda_{\rm USJ}$ is replaced
by $2\lambda_{\rm UJ}/(3-q)$. The true GRB rate for the UJ model
is higher than that for the USJ model for the same fit to the
observed $\log N - \log S$ distribution by a factor of $f(q)$
which is given in Eq. (\ref{f}) and shown in Figure
\ref{fig_f(q)}. In Table \ref{table1} we provide the values for
$q=1$ as an example. The true GRB rate that we obtain for $q=1$ is
a factor of $\approx 6-6.6$ lower than that of \citet{GPW04} and a
factor of $\approx 45-50$ smaller than that of \citet{Frail01}.
However, since the GRB rate that is obtained in this way has a
strong dependence on the value of $q$ (see Figure \ref{fig_f(q)}),
the rates for $q=1$ are not very meaningful (as discussed in the
previous paragraph). A constant efficiency ($b=0$) and a constant
energy ($a=2$) imply $\lambda_{\rm UJ}=2$ which, together with the
best fit value of $\lambda_{\rm USJ}\approx 2.9$, give
$q=3-2\lambda_{\rm UJ}/\lambda_{\rm USJ}\approx 1.6$. For the best
fit parameter values this would imply $f(q)\approx 14.4$ and a
true GRB rate of $R_{\rm GRB}(z=0)\approx 12.3\;{\rm
Gpc^{-3}\;yr^{-1}}$.

\begin{figure}
\plotone{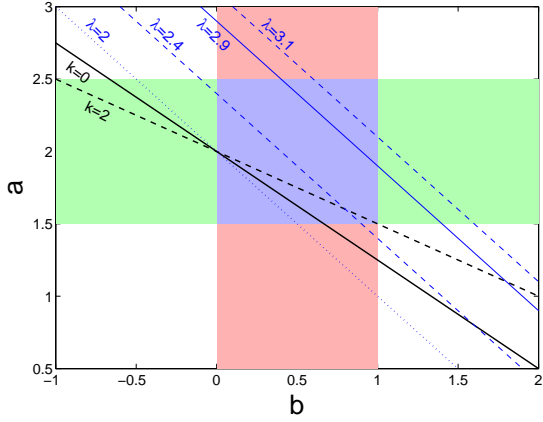} \caption{\label{ab}A summary of the constraints on
the power law indexes $a$ and $b$  of the kinetic energy per solid
angle and the gamma-ray efficiency, respectively:
$\epsilon\propto\theta^{-a}$ and
$\epsilon_\gamma\propto\theta^{-b}$, where $\theta=\theta_{obs}$ for
the USJ model and $\theta=\theta_j$ for the UJ model. The constraint
from the Frail et al. relation (Eq. \ref{lambda}) is shown both for
a constant density external medium ($k=0$) and for a stellar wind
environment ($k=2$). Also shown are the values of $\lambda=a+b$ for
the best fit value ($\lambda_{\rm USJ}=2.9$) and the $1\;\sigma$
confidence interval ($2.4<\lambda_{\rm USJ}<3.1$) for the USJ model,
as well as $\lambda=2$ which is still acceptable. The shaded regions
show the constraints $1.5\lesssim a_{\rm USJ}\lesssim 2.5$ from the
afterglow light curves, which applies only to the USJ model, and
$0\lesssim b\lesssim 1$ from estimates of the gamma-ray efficiency,
which applies to both the USJ model and the UJ model. The
intersection of these two shaded regions is indicated.}
\end{figure}

It is remarkable that we obtain a good fit to the data for the UJ
model for a single power-law distribution of jet half-opening
angles, $P(\theta_j)$, and a scatter $\sigma$ in the peak luminosity
for a given $\theta_j$ which is consistent with observations.
Previous works used a broken power-law luminosity function $P(L)$
which corresponds to a broken power-law $P(\theta_j)$, and no
scatter ($\sigma=0$). For this reason, we also performed a fit with
$\sigma=0$ and a broken power power-law luminosity function $P(L)$
with the parameterization of \citet{GPW04}, where\footnote{Here we
define $\alpha$ and $\beta$ with a minus sign with respect to their
definition in \citet{GPW04}.} $P(L)\propto L^{-1-\alpha}$ for
$L<L_*$ and $P(L)\propto L^{-1-\beta}$ for $L>L_*$. The results of
this fit are shown in Table \ref{table1} and Figures
\ref{cumulative} and \ref{differential}. We get a good fit where
$\alpha=0.6\pm 0.1$ and $\beta=0.8_{-0.1}^{+0.2}$ are consistent
with having the same value, $\alpha\approx\beta\approx 0.7$. This
last value implies $\eta\approx 1.7$, which is consistent with our
best fit value of $\lambda_{\rm USJ}=2.9$ that implies
$\eta=1+2/\lambda_{\rm USJ}\approx 1.7$. This suggests that a broken
power-law is not really necessary, and a single power-law can
provide a reasonable fit to the data even for $\sigma=0$ (as we
obtain for the first fit from the previous section, with different
$L_{\rm min}$ and $L_{\rm max}$ which correspond to different
$\theta_{\rm max}$ and $\theta_c$).

\section{Discussion}
\label{discussion}

We have developed a formalism that enables us to calculate the
theoretical $\log N-\log S$ distribution for any GRB jet
structure, and directly provides the true GRB rate from a fit to
the observed $\log N-\log S$ distribution. This is a more
straightforward approach compared to previous works
\citep{Schmidt01,GPW04} which first calculated the GRB rate
assuming an isotropic emission and then introduced a `correction
factor' in order to account for the effects of the GRB jet
structure. We have applied this formalism to the uniform jet (UJ)
model and to the universal structured jet (USJ) model and
performed fits to the GUSBAD catalog which includes 2204 BATSE
bursts. Our analysis improves on previous works by: (1) allowing
the efficiency in producing $\gamma$-rays to vary with $\theta$,
(2) including an internal dispersion in the peak luminosity $L$ at
any given $\theta$, (3) introducing an inner core angle $\theta_c$
and an outer edge $\theta_{\rm max}$ in the USJ model, and (4)
using a larger and more uniform GRB sample. The results of our
fits are summarized in Table \ref{table1} and Figures
\ref{cumulative} and \ref{differential}.

A power law luminosity function has also been fitted to the the $\log N
- \log S$ distribution in some previous works
\citep[][hereafter STS02]{Hakkila96,LW98,STS02}.  Of these works, it is most useful to
compare our results to those of STS02, since their assumptions are the
closest to ours: similar to us, they assumed
$(\Omega_{\rm M},\Omega_\Lambda)=(0.3,0.7)$ and that the GRB rate follows
the star formation rate.  It is important to keep in mind, however,
that STS02 used a different GRB sample. STS02 obtained a power law
luminosity function at low luminosities, with an index of $\eta=1.4$,
where $P(L)\propto L^{-\eta}$.  This is rather close to the power law
index of $\eta=1.7$ (or between 1.6 and 1.8 at $1\;\sigma$) that we
obtain over the whole luminosity range for a single power law
luminosity function. It is also close to the power law index of
$\eta=1.6$ (or between 1.5 and 1.7 at $1\;\sigma$) that we obtain for
low luminosities when using a broken power law luminosity
function. Also, for a single power law, we have a maximal luminosity
(corresponding to the core angle $\theta_c$ in the USJ model or the
minimal jet half-opening angle $\theta_{\rm min}$ in the UJ model)
with and exponential tail [due to the assumed scatter around the mean
value of $L(\theta)$], which is not very different from an exponential
cutoff (or decline) at high luminosities, that was found to be a
viable option by STS02. Thus, we conclude that our results are
consistent with those of STS02.

The main differences between our work and previous works in the
literature which aimed constraining the GRB luminosity function
using the peak flux distribution are: (i) the different sample that
we use, and (ii) the fact that we consider a differential peak flux
distribution instead of the cumulative distribution, since random
errors propagate in an unknown way in the cumulative distribution.
Moreover, differently from previous authors, we consider the
possibility that the jet is structured and derive the corresponding
luminosity function which in turn puts constraints on the jet
structure through the fit to the $\log N - \log S$ distribution.

The values we obtain for $\theta_c$ in the USJ model or $\theta_{\rm
min}$ in the UJ model are close to the upper limit of $\sim 0.05$ that
is implied by the smallest values of $\theta$ that are inferred from
afterglow observations. Furthermore, they are not consistent with
zero. The value of $\theta_{\rm max}$ that we obtain is close to the
lower limit of $\sim 0.5$ from afterglow observations, and is not
consistent with $\pi/2$. Therefore, we find that including $\theta_c$
(or $\theta_{\rm min}$) and $\theta_{\rm max}$ is required in order to
obtain a good fit to the data.

We fit the observed distribution of $\theta^\lambda L(\theta)$
(where $L$ is the peak luminosity), for a sample of 19 GRBs with a
known redshift $z$ and an estimate for $\theta$, to a log-normal
distribution with a standard deviation $\sigma$. We found the fit
to a log-normal distribution to be acceptable. The implied values
of $\sigma$ are
$(1.0,\,0.9,\,0.8,\,0.9)\lesssim\sigma\lesssim(1.2,\,1.2,\,1.3,\,1.5)$ for
$\lambda=(1.5,\,2,\,2.5,\,3)$, where we have taken into account that
part of the observed dispersion might arise due to error in the
estimated value of $\theta$. This is consistent with the wide
range of possible $\sigma$ values we obtain in our fit for a free
$\lambda$, $\sigma=0.5_{-0.5}^{+0.7}$, and with the smaller range
of values for our fit with a fixed $\lambda_{\rm USJ}=2$,
$\sigma=0.8_{-0.2}^{+0.4}$.

The \citet{Frail01} relation constrains the values of the power law
indexes $a$ and $b$ of the energy per solid angle in the jet
($\epsilon$) and the gamma-ray efficiency ($\epsilon_\gamma$),
respectively (see Eq. \ref{lambda}). For the USJ model the observed
shapes of the afterglow light curves imply $1.5\lesssim a_{\rm
USJ}\lesssim 2.5$ \citep{GK03,KG03}, which in turn implies $11/6
\lesssim \lambda_{\rm USJ} \lesssim 13/6$ and $-2/3 \lesssim b_{\rm
USJ}\lesssim 2/3$ for $k=0$ or $1.5 \lesssim \lambda_{\rm USJ}
\lesssim 2.5$ and $-1\lesssim b_{\rm USJ}\lesssim 1$ for $k=2$. As
discussed in \S \ref{efficiency}, direct estimates of
$\epsilon_\gamma$ from afterglow observations suggest $0\lesssim
b\lesssim 1$. Our best fit value of $\lambda_{\rm
USJ}=2.9^{+0.2}_{-0.5}$ favor a stellar wind external density profile
($k=2$) and an efficiency that decreases with $\theta$
($b>0$). However, $\lambda_{\rm USJ}=2$, which corresponds to $b=0$
and does not constrain the value of $k$, still provides an acceptable
fit to the data. The constraints on $a$ and $b$ are summarized in
Fig. \ref{ab}.

For the UJ model we find a degeneracy in the luminosity function:
$P(L)\propto L^{-\eta}$ where $\eta=1+(3-q)/\lambda_{\rm UJ}$ and
$\lambda_{\rm UJ}=a_{\rm UJ}+b_{\rm UJ}$. Since a fit to the
observed $\log N-\log S$ distribution can only constrain the
luminosity function $P(L)$, in our case it can only provide the
value of $\eta$. However, this constrains only the value of
$(3-q)/\lambda_{\rm UJ}=(3-q)/(a_{\rm UJ}+b_{\rm UJ})$, and
therefore still does not enable us to determine the values $q$ and
$\lambda_{\rm UJ}$ separately. An independent estimate of the true
GRB rate would constrain $f(q)$ and therefore $q$, and would thus
enable one to break this degeneracy.

Alternatively, one can make an additional assumption on $a_{\rm
UJ}$ or $b_{\rm UJ}$. The true energy in the jet scales as
$E\propto\theta_j^{2-a_{\rm UJ}}$ and a constant energy
corresponds to $a_{\rm UJ}=2$. If we assume a constant energy, the
Frail relation (Eq. \ref{lambda}) implies a constant efficiency
($b_{\rm UJ}=0$), and vice versa. In this case $\lambda_{\rm
UJ}=2$ and the best fit value of $\eta=1+2/\lambda_{\rm
USJ}\approx 1.7$ implies $q\approx 1.6$. For the best fit
parameter values this would imply $f(q)\approx 14.4$ and a true
GRB rate of $R_{\rm GRB}(z=0)\approx 12.3\;{\rm
Gpc^{-3}\;yr^{-1}}$. This last value is a factor of $\approx 2.7$
lower than that of \citet{GPW04} and a factor of $\approx 20$
smaller than that of \citet{Frail01}. It is also a factor of
$\approx 4.9\times 10^3$ smaller than the rate of SNe Type Ib/c,
$R_{\rm SN\,Ib/c}(z=0)\approx 6\times 10^4\;{\rm
Gpc^{-3}\;yr^{-1}}$.

For $a_{\rm UJ}=2$ and $b_{\rm UJ}=0$, the $1\;\sigma$ confidence
interval of $2.4<\lambda_{\rm USJ}<3.1$ in $\lambda_{\rm USJ}$
corresponds to $1.3<q<1.7$, $9.5<f(q)<16.5$, and $R_{\rm
GRB}(z=0)\cong (7.7-16.5)\;{\rm Gpc^{-3}\;yr^{-1}}$. These values
of $q$ are not very far from $q=1$ for which there is an equal
number of jets per logarithmic interval in $\theta_j$. It implies
that there are more jets with $\theta_j\sim\theta_{\rm min}$
compared to $\theta_j\sim\theta_{\rm max}$, by a factor of
$\sim(\theta_{\rm max}/\theta_{\rm min})^{q-1}\sim 2.1-9.1$ where
$9\lesssim\theta_{\rm max}/\theta_{\rm min}\lesssim 23$ (see Table
\ref{table1}). If we allow the value of either $a_{\rm UJ}$ or
$b_{\rm UJ}$ to vary, they must both vary together in order to
satisfy the Frail relation (Eq. \ref{lambda}). This would cause
the values of $q$, $f(q)$ and the true GRB rate for the UJ model
to all vary accordingly. The strong dependence of $f(q)$ on $q$
(see Figure \ref{fig_f(q)}) implies that this can potentially
increase the true GRB rate by a large factor. The fact that
$f(q)>1$ for all $q$ implies that the true GRB rate for the USJ
model, which is given in Table \ref{table1}, provides a lower
limit for the UJ model.

\acknowledgements

We thank the referee for useful comments which helped improve the
paper.  D.~G. thanks the Weizmann institute for the pleasant
hospitality.  This work was supported by the RTN ``GRBs - Enigma and
a Tool" and by an ISF grant for a Israeli Center for High Energy
Astrophysics (D.~G.), by the W.~M. Keck foundation and by US
Department of Energy under contract DE-AC03-76SF00515 (J.~G.) and
NSF grants AST-0307502 (D.~G. and M.~C.~B.) and PHY-0070928 (J.~G.).

\end{document}